\documentclass[preprint,11pt]{aastex}

\usepackage{emulateapj5}

\newcommand{\eten}[1]{\mbox{$10^{#1}$}}

\newcommand{\degree}{\mbox{$^{\circ}$}}

\newcommand{\um}{$\mu$m}

\newcommand{\iras}{\mbox{\it IRAS}}
\newcommand{\iso}{\mbox{\it ISO}}

\newcommand{\sirtf}{\mbox{\it SIRTF}}
\newcommand{\SIM}{\mbox{\it SIM}}
\newcommand{\herschel}{\mbox{\it Herschel}}
\newcommand{\ngst}{\mbox{\it JWST}}
\newcommand{\sofia}{\mbox{\it SOFIA}}
\newcommand{\sma}{\mbox{SMA}}
\newcommand{\alma}{\mbox{ALMA}}
\newcommand{\carma}{\mbox{CARMA}}
\newcommand{\tpf}{\mbox{\it TPF/Darwin}}
\newcommand{\hst}{\mbox{\it HST}}

\newcommand{\irac}{\mbox{\rm IRAC}}
\newcommand{\mips}{\mbox{\rm MIPS}}
\newcommand{\irs}{\mbox{\rm IRS}}
\newcommand{\ssc}{\mbox{\rm SSC}}
\newcommand{\gto}{\mbox{\rm GTO}}
\newcommand{\feps}{\mbox{\rm FEPS}}

\newcommand{\simba}{\mbox{\rm SIMBA}}
\newcommand{\sest}{\mbox{\rm SEST}}
\newcommand{\scuba}{\mbox{\rm SCUBA}}
\newcommand{\jcmt}{\mbox{\rm JCMT}}
\newcommand{\cso}{\mbox{\rm Caltech Submillimeter Observatory}}
\newcommand{\mambo}{\mbox{\rm MAMBO}}
\newcommand{\complete}{\mbox{\rm COMPLETE}}
\newcommand{\bolocam}{\mbox{\rm Bolocam}}
\newcommand{\iram}{\mbox{\rm IRAM}}
\newcommand{\mm}{millimeter}
\newcommand\submm{submillimeter}

\newcommand\sed{spectral energy distribution}
\newcommand{\lsun}{\mbox{L$_\odot$}}
\newcommand{\msun}{\mbox{M$_\odot$}}
\newcommand{\mmoon}{\mbox{M$_{Moon}$}}

\newcommand{\lbol}{\mbox{$L_{bol}$}} 

\newcommand{\av}{\mbox{$A_V$}} 

\newcommand{\cooo}{C$^{18}$O}

\newcommand{\hcop}{HCO$^+$}

\newcommand{\nnhp}{N$_2$H$^+$}

\newcommand{\jj}[2]{\mbox{$J = #1\rightarrow#2$}}

\input{epsf}

\def\plotfiddle#1#2#3#4#5#6#7{\centering \leavevmode
\vbox to#2{\rule{0pt}{#2}}
\includegraphics{#1}}

\newcommand{\mjup}{M$_{Jup}$}
\newcommand{\mearth}{M$_{Earth}$}
\newcommand{\pms}{pre-main-sequence}
\newcommand{\tnm}[1]{\mbox{\tablenotemark{#1}}}

\slugcomment{\footnotesize {}}
\shorttitle{From Cores to Disks}
\shortauthors{Evans et al. }

\begin{document}


\title {\bf From Molecular Cores to Planet-forming Disks: A \sirtf\
Legacy Program }
\author {Neal J. Evans II}
\affil{ The University of Texas at Austin, Department of Astronomy, 
       1 University Station C1400, Austin, Texas 78712--0259}
\email{nje@astro.as.utexas.edu}
\author{Lori E. Allen}
\affil{Smithsonian Astrophysical Observatory, 60 Garden St. MS42, 
Cambridge, MA 02138}
\email{leallen@cfa.harvard.edu}
\author{Geoffrey A. Blake}
\affil{Division of Geological and Planetary Sciences 150-21, 
California Institute of Technology, Pasadena, CA 91125}
\email{gab@gps.caltech.edu}
\author{A. C. A. Boogert}
\affil{Division of Physics, Mathematics, \& Astronomy 105-24, 
California Institute of Technology , Pasadena CA 91125}
\email{acab@astro.caltech.edu}
\author{Tyler Bourke}
\affil{Smithsonian Astrophysical Observatory, 60 Garden St. MS42, 
Cambridge, MA 02138}
\email{tbourke@cfa.harvard.edu}
\author{Paul M. Harvey}
\affil{ The University of Texas at Austin, Department of Astronomy, 
       1 University Station C1400, Austin, Texas 78712--0259}
\email{pmh@astro.as.utexas.edu}
\author{J. E. Kessler}
\affil{Division of Chemistry \& Chemical Engineering, California Institute of
Technology, Pasadena, CA 91125}
\email{kessler@its.caltech.edu}
\author{David W. Koerner}
\affil{Northern Arizona University, Department of Physics and Astronomy, 
Box 6010, Flagstaff, AZ 86011-6010}
\email{koerner@physics.nau.edu}
\author{Chang Won Lee}
\affil{Taeduk Radio Astronomy Observatory, Korea Astronomy Observatory,
36-1 Hwaam-dong, Yusung-gu, Taejon 305-348, Korea}
\email{cwl@trao.re.kr}
\author{Lee G. Mundy  }
\affil{Astronomy Department, University of Maryland, College Park, MD 20742}
\email{lgm@astro.umd.edu}
\author{Philip C. Myers}
\affil{Smithsonian Astrophysical Observatory, 60 Garden St. MS42, 
Cambridge, MA 02138}
\email{pmyers@cfa.harvard.edu}
\author{Deborah L. Padgett}
\affil{SIRTF Science Center, MC 220-6, Pasadena, CA 91125}
\email{dlp@ipac.caltech.edu} 
\author{K. Pontoppidan}
\affil{Leiden Observatory, Postbus 9513, 2300 RA Leiden, Netherlands}
\email{pontoppi@strw.leidenuniv.nl}
\author{Anneila I. Sargent}
\affil{Division of Physics, Mathematics, \& Astronomy 105-24, 
California Institute of Technology , Pasadena CA 91125}
\email{afs@astro.caltech.edu}
\author{Karl R. Stapelfeldt}
\affil{MS 183-900, Jet Propulsion Laboratory, California Institute of 
Technology, 4800 Oak Grove Drive, Pasadena, CA 91109}
\email{krs@exoplanet.jpl.nasa.gov} 
\author{Ewine F. van Dishoeck}
\affil{Leiden Observatory, Postbus 9513, 2300 RA Leiden, Netherlands}
\email{ewine@strw.leidenuniv.nl}
\author{Chadwick H. Young}
\affil{ The University of Texas at Austin, Department of Astronomy, 
       1 University Station C1400, Austin, Texas 78712--0259}
\email{cyoung@astro.as.utexas.edu}
\author{Kaisa E. Young}
\affil{ The University of Texas at Austin, Department of Astronomy, 
       1 University Station C1400, Austin, Texas 78712--0259}
\email{kaisa@astro.as.utexas.edu}

\begin{abstract}

Crucial steps in the formation of stars and planets can be studied
only at mid-infrared to far-infrared wavelengths, where \sirtf\ provides
an unprecedented improvement in sensitivity. We will use all three \sirtf\
instruments (\irac, \mips, and \irs ) to observe sources
that span the evolutionary sequence from molecular cores to protoplanetary
disks, encompassing a wide range of cloud masses, stellar masses, and
star-forming environments. In addition to targeting about 150 known compact
cores, we will survey with \irac\ and \mips\ (3.6 to 70 \micron)
the entire areas of five of the nearest
large molecular clouds for new candidate protostars and substellar objects
as faint as 0.001 solar luminosities. We will also observe with \irac\ and 
\mips\
about 190 systems likely to be in the early stages of planetary system formation
(ages up to about 10 Myr), probing the evolution of the circumstellar dust,
the raw material for planetary cores.
Candidate planet-forming disks as small as 0.1 lunar masses will be
detectable.  Spectroscopy with \irs\ of new objects found in the surveys and of
a select group of known objects will add vital information
on the changing chemical and physical conditions in the disks and envelopes.
The resulting data products will include catalogs of thousands of
previously unknown sources, multiwavelength maps of about 20 square
degrees of molecular clouds, photometry of about 190 known young stars,
spectra of at least 170 sources, ancillary data from ground-based
telescopes, and new tools for analysis and modeling.
These products will constitute the foundations for many
follow-up studies with ground-based telescopes, as well as with
\sirtf\ itself and other space missions such as \SIM, \ngst, \herschel, and 
\tpf.

\end{abstract}


\section{Introduction}

Stars and planets form in a closely coupled process that is generally
accessible to study only at relatively long wavelengths, from infrared
to radio. Observational studies of this process have generally suffered
from one or more of the following problems: biased samples,
inadequate sensitivity, inadequate spatial resolution, or incomplete data 
across the wavelength ranges of interest.
The {\it Space Infrared Telescope Facility} (\sirtf) offers a singular
opportunity for a major advance in this area of research. The \sirtf\
mission has been described by Gallagher, Irace, \& Werner (2002).

Our Legacy program, ``From Molecular Cores to Planet-forming Disks,"
uses 400 hours of \sirtf\ observations 
to study the process of star and planet formation
from the earliest stages of molecular cores to the epoch of planet-forming
disks. This program, hereafter referred to simply as ``Cores to Disks" or
c2d, is closely coordinated with the Legacy program on ``Formation and
Evolution of Planetary Systems" or FEPS, which carries the age sequence to
later times. Our program uses all three \sirtf\ instruments: 
the {\it InfraRed Array Camera} or \irac, covering 3.6 to 8 \micron\ (Fazio
et al. 1998);
the {\it Multiband Imaging Photometer for SIRTF} or \mips, covering 24 to
160 \micron\ (Engelbracht et al. 2000);
and
the {\it InfraRed Spectrometer}, or \irs, supplying spectroscopy from 
5.3 to 40 \micron\ with resolving power $R = 60-120$ and from 
10 to 37 \micron\ with $R = 600$ (Houck et al. 2000).

The observational programs, described in greater detail in the individual
sections, include unbiased mapping of 5 large nearby molecular clouds
and about 150 compact molecular cores (\S \ref{cores}), 
photometry of about 190 stars with ages up to about 10 Myr (\S \ref{stars}), 
and spectroscopy of at least 170 objects in a wide range of evolutionary states
(\S \ref{irs}). In these sections, we discuss the scientific questions,
the sample selection, the planned observations, and the expected results. 
The data products are summarized in \S \ref{datprod}.
We also describe {\it ancillary} data, which are part of the
c2d data products, and {\it complementary} data, which are being obtained by
us or others to provide a more complete picture.
The \sirtf\ and ancillary data will be available to the broader community
from the {\it SIRTF Science Center} (\ssc), via their Infrared Sky
Archive (IRSA).
Complementary data products will be made available, as far as possible, 
through either IRSA or public web sites.
Further information on the program can be found at the 
c2d website: {\tt http://peggysue.as.utexas.edu/SIRTF/}.


\section{A Survey of Nearby Molecular Clouds and Cores} \label{cores}

\subsection{Scientific Questions}

The \sirtf\ mission offers an unprecedented opportunity to determine 
the stellar content of the nearest star-forming molecular clouds, the 
distributions of their youngest stars and substellar objects, and the 
properties of their circumstellar envelopes and disks.  Just as \iras\ and 
\iso\ revealed many properties of isolated and clustered star-forming 
regions, \sirtf\ will yield new insights into how stars and brown dwarfs
are born.  Among the specific questions that our program 
will address are the following.

1. In large cloud complexes, how are the youngest stars and 
substellar objects distributed, in position and mass?  
The \irac\ and \mips\ maps of large complexes will reveal the 
reddest, and presumably youngest, associated objects. Is their 
distribution ``bimodal" between singles and clusters, or is there a 
more continuous distibution of multiplicity?  Is the distribution of 
young stars better correlated with line-of-sight extinction, the 
supply of dense gas, or with proximity to other young stars, a 
measure of triggering or cooperative star formation?  Do the answers 
to these questions depend on the mass of the object -- for example, 
can brown dwarfs form in isolation, as can ordinary low-mass stars, 
or do they require proximity to stars, as expected if they originate 
primarily in circumstellar disks or in small stellar groups?

2. What is the incidence of circumstellar disks in complexes and in 
isolated cores?  The excess emission over that of a stellar 
photosphere at near- and mid-infrared wavelengths indicates the 
presence of a circumstellar disk, the birthplace of planets. To 
improve understanding of the factors that influence disk dissipation and 
survival, our sample includes a wide range of
complex and isolated star-forming regions.  This 
comparison will differ from past studies of disk incidence in its 
greater sensitivity to faint objects, in its unbiased coverage of 
large cloud areas, and in its inclusion of isolated cores not in 
complexes.

3. Do ``starless" isolated cores harbor faint protostars?  Starless 
or ``pre-protostellar" cores have no pointlike infrared emission 
according to ground-based near-infrared observations and 
far-infrared observations by the \iras\ and \iso\ satellites.  These 
cores are prime targets for studies of the initial conditions of 
isolated star formation, and some such cores have associated internal motions 
suggestive of the early stages of star formation.  \sirtf's 
sensitivity will allow detection of extremely young protostars and
proto-brown-dwarfs missed 
by earlier observations.  Identification of such objects would 
stimulate more detailed studies and so would advance our 
understanding of the earliest stage of star formation.

4. What are the statistical lifetimes of various stages?
Our survey covers a large area and will catalog a large number
of young and embedded systems. These data will allow the first
unbiased statistical studies of the complete stellar content of
five large clouds. In addition, systems that are statistically
rare and systems in rapid phases of evolution will be present in
the sample. We may, for example, find systems in the
short-lived phase in which the first hydrostatic core forms, a crucial, 
but so far unobserved stage of star formation (Boss \& Yorke 1995).

5. Do isolated cores with associated stars preferentially harbor 
single stars or small stellar groups?  Observations by \iras\ and \iso\ 
led to the idea of ``one core, one star": an isolated core or 
globule tends to form a single star or an unresolved binary, as 
opposed to a small group of 3 to 10 members. In large molecular clouds, 
many isolated \iras\ point sources are accompanied by groups of 
near-infrared sources that constitute the more evolved members of a 
small stellar group (e.g., Hodapp 1994). \sirtf\ will be much more 
sensitive to faint emission from nearby sources than was \iras\ and 
\iso, allowing us to test whether truly isolated star formation
occurs.

6. What is the density structure of individual cores? Some isolated 
starless cores have a density structure similar to that of a 
pressure-bounded isothermal sphere, according to observations of 
their \submm\ dust emission and of their near-infrared absorption of 
background starlight.  These cores have a single local maximum of 
density and tend to make single stars.  Line profiles toward these cores
usually indicate low levels of turbulence.  In contrast, cores 
in cluster-forming regions appear to have more complex density 
structure, and their spectral lines indicate more turbulent conditions.  
Deep \irac\ imaging of such cores will reveal their detailed 
structure through mapping the extinction of background stars.  This 
structure will be compared with that of isolated cores to improve
 understanding of the initial conditions for stellar groups.

\subsection{Sample Selection}

Our sample of the nearest star-forming molecular clouds covers
a range of cloud ``types'' broad enough to encompass all
modes of star formation and sufficient in number to allow robust
statistical conclusions.
To ensure a wide variety of star-forming conditions, we target five large
complexes known to be forming stars in isolation, in groups, and in
clusters, and 156 small, isolated cores -- 110 starless and 46 with associated
\iras\ sources. 

The five large clouds selected for mapping, listed in Table 1, 
satisfy the following criteria: 
they contain a substantial mass of molecular gas; 
they are actively forming stars;
they are within 350 pc; and they can be mapped in a 
reasonable amount of time.
Table 1 gives for each target region the distance, the area to be
mapped with IRAC, and the full time for IRAC and MIPS observations,
including off-cloud comparison fields.
Although the distance criterion excludes some important star-forming 
regions, such as the Orion clouds, it was imposed to ensure that our 
\sirtf\ observations would be sensitive to brown dwarfs. 
The Perseus cloud is the most distant cloud at 320 pc (de Zeeuw et al.
1999), though previous distance estimates range from 220 pc (\v{C}ernis 1990)
to 350 pc (Herbig \& Jones 1983).
The Taurus molecular cloud satisfies most criteria but is too large 
to be mapped within the time allocated.

The selected complexes span a wide range of conditions.
The most quiescent and least opaque are Lupus
and Chamaeleon II, whose young stars are isolated or in sparse
groups, and whose gas has low extinction and narrow lines.  The most
turbulent complexes are Perseus, Ophiuchus, and Serpens, with more densely
clustered stars. The guaranteed time observers (\gto s) will observe smaller
regions in many of these clouds. Our goal is to provide complete
and unbiased coverage down to a set \av\ limit.

Selection of the compact molecular cores proceeded in three steps:
selection of a large sample with inclusive criteria; classification into
cores with and without known internal luminosity sources (``stars"); and
reduction of the sample to eliminate objects being observed by \gto s and
to fit within the allowed time. The latter requirement resulted in removing
cores with less robust evidence for dense gas and dust.
The initial criteria for selection of the isolated dense core sample were 
that the distance is less than 400 pc, that the size is not too large
(typically less than $5\arcmin$), and 
that the core has been
mapped in a dense gas tracer, typically NH$_3$ (Jijina, Myers \& Adams
1999), CS (Lee, Myers \& Tafalla 2001), and/or \nnhp\ (Lee et al.
2001; Caselli et al. 2002).  To supplement this list, we searched
the ADS and astro-ph databases for other cores mapped in these tracers but
not included in these catalogs, or mapped in
other dense gas tracers, such as HCO$^+$, HCN, and H$_2$CO.  
We also added cores with dust continuum  emission but
without published maps in tracers of dense gas.
Cores where more than one dense gas tracer had been
detected, but no map exists, were added to include Bok globules whose
size can be estimated from their optical extinction.

To determine whether a core has an associated
``star," meaning a central luminosity source at any evolutionary stage,
we used the catalog of Lee \& Myers (1999).  
For embedded sources (Class 0 and Class I), they
searched the IRAS database and required the following to be true.
The projected position of the IRAS source on the sky should
fall within the contour of least extinction defining the optical
extent of the core.
The source should be detected in at least two wavebands.
The detected flux densities ($F$) should be greater in the longer
wavelength band.  However, sources with $F_{100\mu m} < F_{60\mu m}$
or $F_{100\mu m} < F_{25\mu m}$ were included as long as
$F_{60\mu m} > F_{25\mu m} > F_{12\mu m}$.
To select \pms\ stars from the IRAS catalog the color-color criteria of
Weintraub (1990) was used by Lee \& Myers (1999), which requires
$-2.00<\log(\nu_{12}F_{12}/\nu_{25}F_{25})/\log(\nu_{12}/\nu_{25})<1.35$ and
$-1.75<\log(\nu_{25}F_{25}/\nu_{60}F_{60})/\log(\nu_{25}/\nu_{60})<2.20$.
The Herbig \& Bell (1988) catalog of \pms\ stars was also searched.
Other cores with stars were identified on a core by core basis from the
literature; e.g., IRAM 04191+1522 was not detected by IRAS, but
was identified by its dust continuum emission (Andr\'e, Motte \& Bacmann
1999) and found to have a central source.

This candidate target list 
was cross-checked against the  \gto\ lists and the area to
be mapped by us in the 5 large molecular clouds to remove
overlap. In order to fit the time allocation,
the list was further reduced by eliminating cores not detected at all 
in CS and \nnhp, or detected, but not mapped, in a dense gas tracer
and detected only weakly in \mm\ or \submm\ continuum.

\subsection{Planned Observations and Expected Results}

The molecular cloud complexes and isolated cores will be observed with
both \irac\ and \mips\ at all wavelengths from 3.6 to 160 \micron,
but we expect the 160 $\mu$m detector to be saturated toward these regions.
Because most of the molecular clouds in our sample
are at low ecliptic latitude ($-40\degree \le \beta \le 40\degree$),
all observations will be made at two 
epochs separated by 3 to 6 hours in order to identify 
faint asteroids. One set of observations will be made in 
the high dynamic range mode of IRAC to minimize saturation on 
bright sources.

\begin{figure}
\plotone{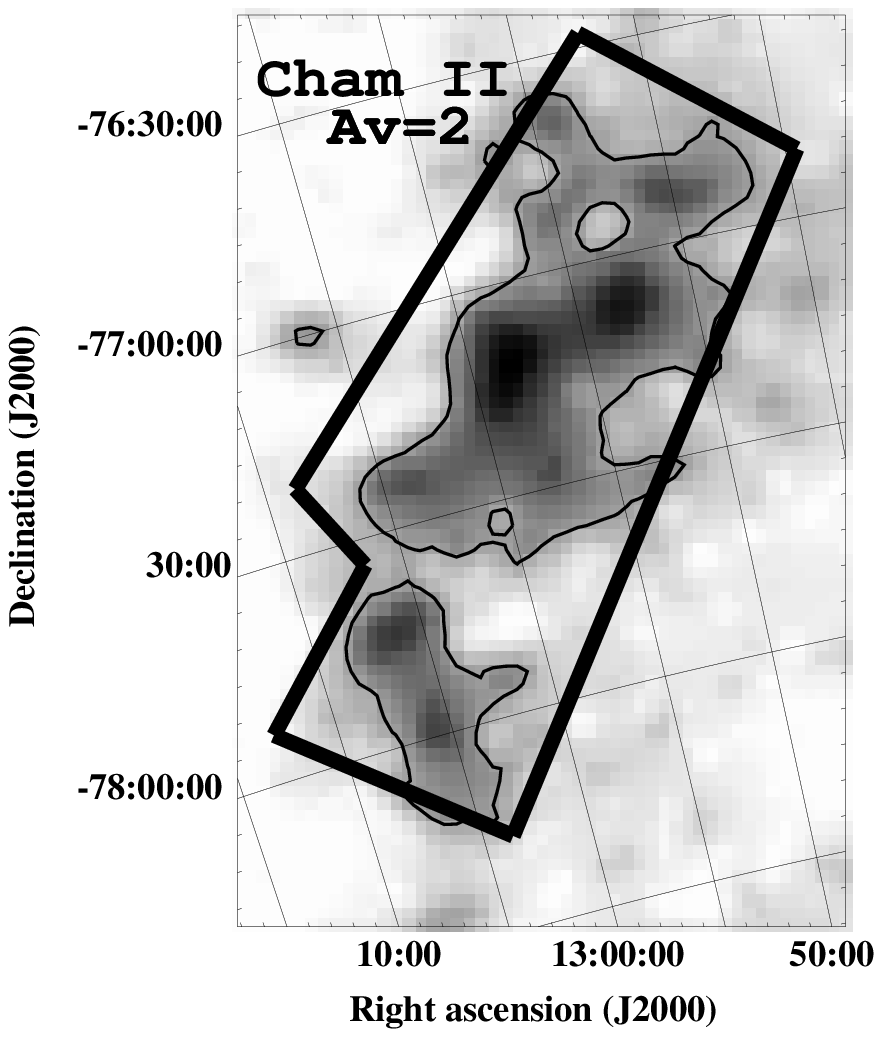}
\figcaption{\label{cham} The observations for Chamaeleon II are outlined on this optical extinction map (Cambr\'{e}sy 1999).
The contour shows $\av =2$ and was used in planning.}
\end{figure}

\begin{figure}
\plotone{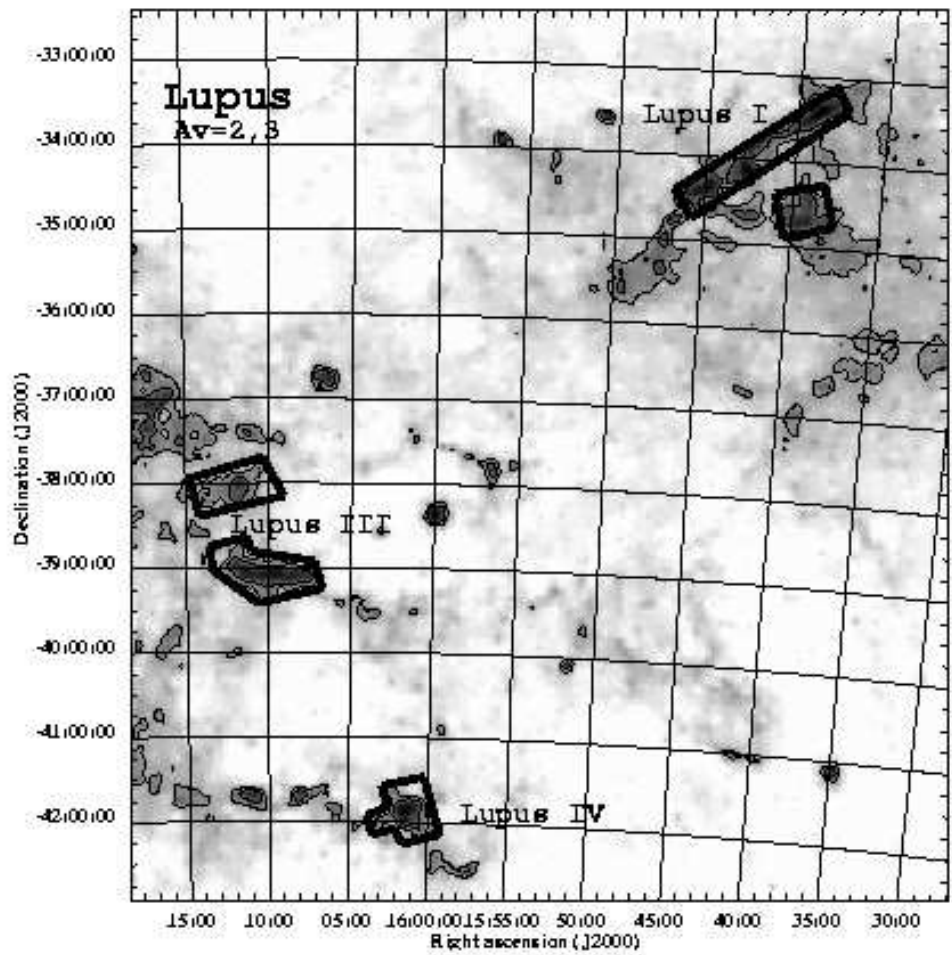}
\figcaption{\label{lupus} We planned observations of Lupus I, III, and IV based on the $\av =2$ contour (for III and IV) 
and the $\av =3$ contour (for Lupus I). The optical extinction map is from 
Cambr\'{e}sy (1999)}
\end{figure}

\begin{figure}
\plotone{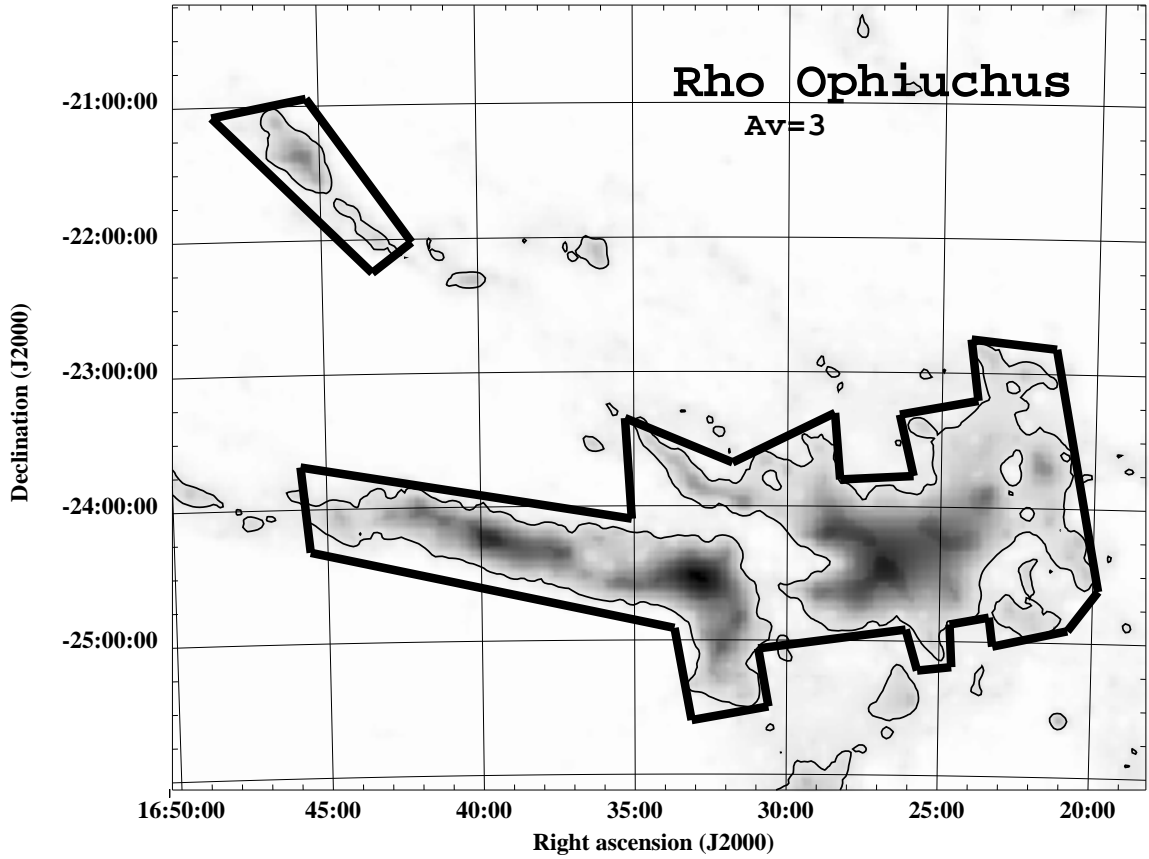}
\figcaption{\label{rhooph} The planned observations for Rho Ophiuchus are outlined around the $\av =3$ contour (Cambr\'{e}sy 1999).}
\end{figure}

\begin{figure}
\plotone{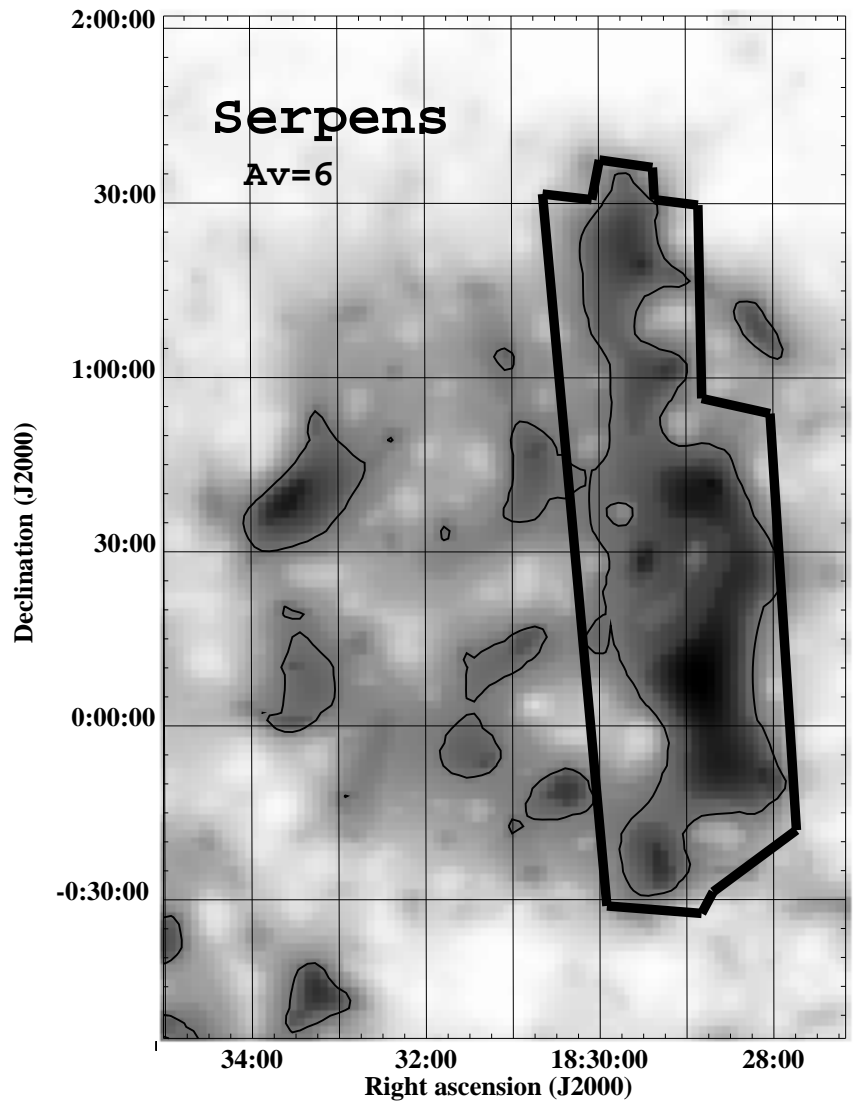}
\figcaption{\label{serpens}The planned observations for Serpens are outlined around the $\av =6$ contour (Cambr\'{e}sy 1999).}
\end{figure}

\begin{figure}
\plotone{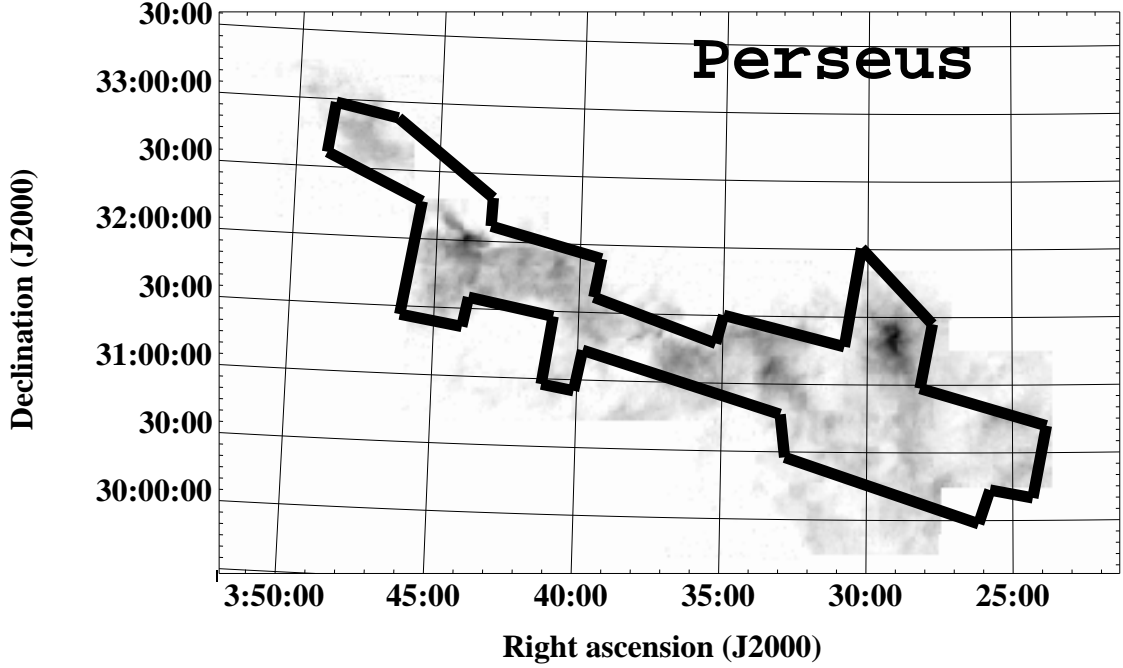}
\figcaption{\label{perseus}Observations of Perseus in $^{13}$CO 
(in grayscale) were 
primarily used in the SIRTF planning for this molecular 
cloud (Padoan et al. 1999).
However, we also used maps of near-infrared colors from the 2MASS Point
Source Catalog (Cutri et al. 2001) to provide constraints similar 
to those for the
other clouds;  the outlined region in this figure corresponds to $\av \sim 2$. }
\end{figure}

The boundaries for four of the five large cloud maps were defined 
using optical extinction 
maps from Cambr\'{e}sy (1999) (see Figures~\ref{cham} to \ref{serpens}). 
A $^{13}$CO map (Padoan et al.\ 1999, Figure~\ref{perseus}) was 
used to define the Perseus molecular cloud. The \sirtf\ map of Perseus 
will include nearly 
all of the area mapped in $^{13}$CO; this region corresponds to an $\av \sim 2$
as inferred from maps of near-infrared colors 
(Carpenter, pers. comm.) generated using the
2MASS Point Source Catalog (Cutri et al. 2001).
For the other clouds, an \av\ level was chosen 
so that each map could cover a continuous area of high extinction
within the allocated time.
In most cases, $\av = 2$ or 3 defined the cloud edge. However, the Serpens 
cloud boundary was limited to $\av  = 6$ because lower levels merge into
unrelated extinction in the Galactic plane. 

Several factors shaped the detailed observing strategy for the large clouds.
In Ophiuchus and Perseus, some regions will saturate the 
\mips\ 70 \micron\ detector. Each saturated region was quarantined by limiting it
to a single \sirtf\ Astronomical Observation Request (AOR) so that the data in the 
surrounding regions would not be adversely affected. Point sources will 
also saturate; the after-image may leave a trail in the direction of the 
\mips\ scans. In order to obtain usable data in such regions, 
\mips\ will scan in the opposite direction for the second epoch observations. 
Additionally, many of our cloud maps overlap
designated \gto\  regions. Therefore, unnecessary duplicate 
observations had to be avoided. Another factor considered in 
the observation planning was cloud orientation. 
The sky orientation of SIRTF maps for high ecliptic latitude targets, 
such as Chamaeleon II and Serpens, rotates rapidly.
In order to eliminate the possibility of gaps in the maps
due to this rotation, the overlap between adjacent AORs was increased, or 
the observations were constrained to specific position angles.

We also plan observations of off-cloud positions in order to sample 
properly the background source counts.  
In the selection of these off-positions, we implemented three criteria.  
First, in order to sample the stellar density gradient as a 
function of galactic latitude, we chose off-cloud positions at different 
heights above the galactic plane. Next, we required that $\av < 0.5$ 
in the off-cloud regions, based on the optical extinction 
maps of Cambr\'{e}sy (1999).  Finally, the off-cloud regions
were required to have little or no molecular emission ($^{12}$CO), based on 
the maps of Dame et al. (2001).
The off-cloud regions are mostly 15\arcmin$\times$15\arcmin.  
Using a Galactic model (Wainscoat et al. 1992), we have 
predicted the stellar background counts for each field (Table 
\ref{background_tab}) and have chosen the size of the off-cloud region 
so that we will detect at least 100 background stellar sources at 3.6 \micron.
The counts in these  off-cloud regions are often much greater than 100---
e.g., at $b=2\degree$ (for Serpens), 
we expect more than $10^5$ background stellar 
counts in the 15\arcmin$\times$15\arcmin\ region.

The cloud maps will be made using slightly overlapping \irac\ frames and
\mips\ scan legs. The \irac\ cloud maps will employ 2 dithers with an
exposure time of 12 seconds in 2 epochs (24 sec total).
Since each cloud will be mapped twice,
one map will be made in \irac's high dynamic range mode (HDR), which
uses exposure times of 12 and 0.6 seconds, the shorter exposure time
enabling photometry of bright point sources that might be saturated in
the 12 second frames. Isolated clouds will be observed with the same
parameters as large cloud complexes; expected sensitivities of the
\irac\ molecular cloud observations are given in Table \ref{senstab}.

The cloud area mapped with \mips\ will be
at least 20\% larger than that mapped by \irac\, because of the 
efficiency of \mips\ scan mapping and the instrument-limited choice of 
scan lengths.
Using the \mips\ fast scan mode, we will obtain 15 sec exposures at each
epoch and achieve the sensitivities given in Table \ref{senstab}.
\mips\ observations of isolated clouds will be made in large-source
photometry mode.
At 24 $\mu$m, an integration time of 3 sec (for a total time, including
2 epochs, of 72 sec) will be used.
At 70 \micron, isolated clouds that contain no \iras\ sources,
or \iras\ sources fainter
than 2 Jy at 60$\mu$m will be observed in the large-source,
coarse-scale mode,
with an integration time of 3 sec (for a total time, including 2 epochs, 
of 36 sec).
Clouds that contain \iras\ sources with $2<S_{\nu}(60\micron)<10$ Jy, will
be observed at 70 \micron\ in the large-source super-resolution mode
(fine scale), with an integration time of 3 sec (for a total time, including
2 epochs, of 48 seconds).

\begin{figure}
\plotfiddle{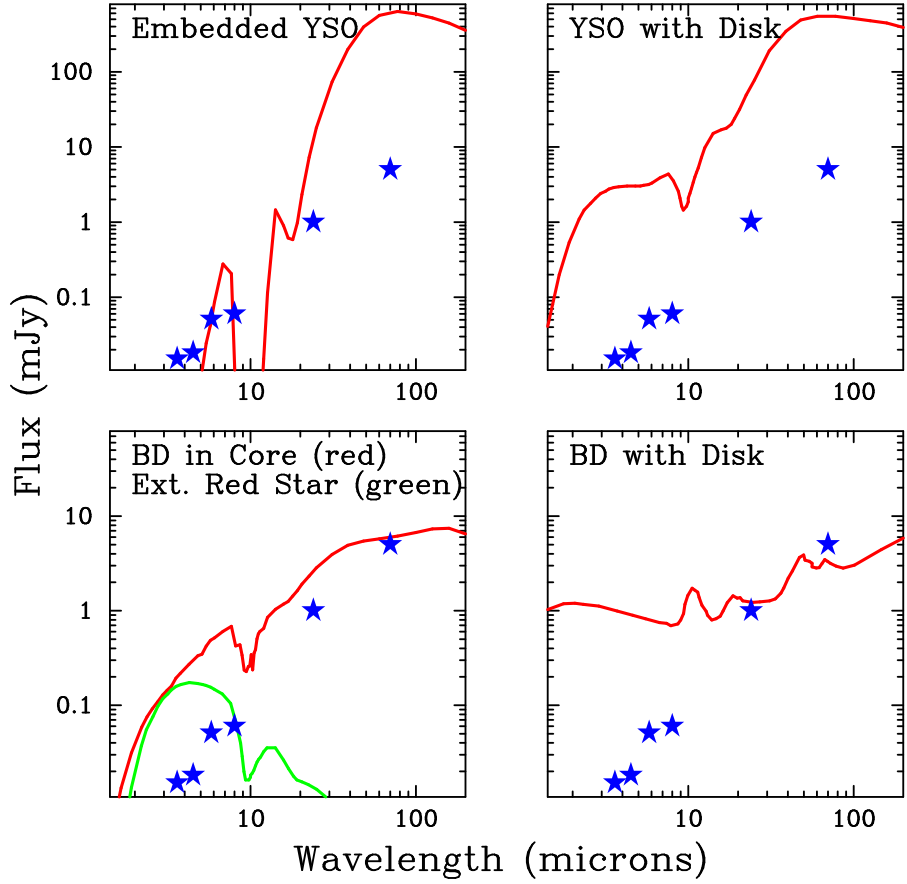}{6.5in}{-90}{65}{65}{-250}{450}
\figcaption{Flux density distributions for sources at 300 pc.
Clockwise from upper left,
a deeply embedded 0.1 \lsun\ protostar, a lightly
embedded 0.1 \lsun\ star with a 30 AU circumstellar disk,
a 10 Myr old, 0.007 \lsun\ brown dwarf with a 4.5 \mjup\ disk, and
a young 0.003 \lsun\ brown dwarf with a 1 \mjup\ envelope
(Chiang et al. 2000, pers. comm.).
The lower curve in the lower left panel shows a background giant star
with the same temperature, and with the same extinction, as the brown dwarf.
The stars represent our 3$\sigma$ IRAC and MIPS sensitivity limits.
\label{seds}
}
\end{figure}

The $3\sigma$ limits from Table \ref{senstab}  for the scan maps yield a limit
on \lbol\ from 3.6 to 70 \micron\ of about \eten{-3} \lsun\ at 350 pc.
Figure \ref{seds} shows (clockwise from upper left) 
the survey detection limits against typical
\sed s for a heavily embedded protostar, a less embedded protostar with a disk,
a brown dwarf with a disk of 4.5 M$_{Jup}$, and an embedded brown dwarf.
Our simulations indicate that disks as small as 1 \mearth\  could be detected, 
even when buried behind $\av \approx 100$ mag of extinction.
The \irac\ bands would detect a 1 Myr old, 5 \mjup\ brown dwarf at 350 pc, 
based on models by Burrows et al. (1997). 
The five large molecular complexes contain about 750
known stars and \iras\ point sources.
Based on recent estimates of the
initial mass function in the field and in young clusters (Meyer et al. 2000, 
Luhman et al. 2000), we expect to identify
several thousand new substellar objects. 
These must be distinguished from a large
number of background stars (Table \ref{background_tab}) and galaxies
(e.g., Lagache et al. 2003) by color criteria (Fig. \ref{seds}) and 
follow-up observations.

\subsection{Ancillary and Complementary Data}

To extend the data base to longer wavelengths, where the dust emission becomes
optically thin and a good tracer of mass, we are obtaining ancillary data at
millimeter wavelengths, using \bolocam\ for those clouds accessible from the
\cso\ (Perseus, Ophiuchus, and Serpens). In addition, we are obtaining
complementary data for the southern clouds (Lupus and Chamaeleon) and isolated
cores using the  Sest IMaging Bolometer Array (\simba) on the 
Swedish-ESO Submillimeter Telescope (\sest). 
Further complementary data on the northern clouds in
molecular lines and dust extinction and emission are being obtained by the 
\complete\footnote{http://cfa-www.harvard.edu/$\sim$ agoodman/research8.html}
team, who also plan to make the data public.

Complementary data on the isolated cores are
also being taken with the  Submillimetre Common User Bolometer Array
(\scuba) on the  James Clerk Maxwell Telescope (\jcmt) and
\mambo\ on the 30-m telescope of \iram. Complementary
spectral line data to extend the CS \jj21 and N$_2$H$^+$ \jj10 observations
of Lee et al. (2001) and Caselli et al. (2002) are being obtained
with the 32 element SEQUOIA focal-plane array on the
Five College Radio Astronomy Observatory (FCRAO) 14-m telescope.
These complementary data,
which will be made available to the community through our c2d website,
provide information on the existence and distribution of cloud material
around objects detected at shorter wavelengths.
Such information will be valuable for establishing
the association of specific SIRTF sources with the cloud, for comparison
of the material and stellar distributions, and for completing the
overall picture of how star formation proceeds in clouds.

Complementary gas and dust observations on smaller
scales are being acquired with the BIMA millimeter wavelength array.
A mapping survey of 21 embedded cores,
mostly in the Perseus complex in the $\lambda = 2.7$ mm continuum and
$^{13}$CO and C$^{18}$O \jj10 lines, has been completed. 
The observed cores include
well-known young embedded systems such as NGC 1333 IRAS4,
NGC 1333 IRAS7, and B5 IRS1, as well as ammonia cores
without known embedded sources. The maps have a resolution of
2\arcsec\ and a field of view of 100\arcsec. This combination
of resolution and field of view will allow detailed comparisons of
the gas, dust, and IRAC-derived stellar distributions. The typical
continuum sensitivity of 1 mJy/beam will permit detection of 
circumstellar disks down to a mass of 0.01 \msun. The
molecular data will also provide velocity structure information,
which can be used to look at the correlation between turbulent/systematic 
velocity fields and stellar binarity and the stellar spatial distribution.

In addition, as complementary data, the large clouds and many northern cores are
being imaged at shorter (R, i, and z) wavelengths to limiting
magnitudes of 24.5, 22, and 22, respectively.  The photometry at these
wavelengths will be very valuable for identifying young objects,
discriminating against background and foreground stars, and extending
the coverage of the \sed. Extended objects will also be resolved, and
young reflection nebulae and (contaminating) background galaxies will
be identified. Data will be obtained primarily with the CFH12K and Megacam
cameras at the Canada-France-Hawaii Telescope. Southern targets (such
as Chamaeleon) will be imaged in similar bands with WFI at ESO, La Silla.


\section{The Evolution of Disks up to 10 Myr} \label{stars}

\subsection{Scientific Questions}

Disks around pre-main sequence stars are the likely sites of planet
formation. As such, they have been targets of extensive research over 
the past two decades.  Surveys of continuum emission
at infrared and millimeter wavelengths have 
established that 50\% of all classical T Tauri
stars (cTTs) (ages $<$ 3 Myr) have disks with typical masses 
of $10^{-3} - 10^{-1}$ M$_{\odot}$, sufficient to form a
planetary system like our own (see Beckwith \& Sargent 1996, Beckwith
1999 for reviews).  Disk sizes, as revealed by millimeter interferometry
and \hst\ images, range from 100 to 1000 AU in diameter.  Much less is known
about the occurrence and properties of disks in later stages of evolution.
Evidence suggests that disk dispersal is marked by the 
termination of viscous accretion onto the star, dissipation of
circumstellar gas, and the coagulation of grains into larger particles.
During this process, disks are thought to evolve from an early opaque
stage to one in which they are optically thin to both stellar radiation
and their own thermal infrared emission.  Examples of such ``debris disks'' 
are found in association with nearby stars at ages up to about 1 Gyr and
contain relatively small amounts of dust (less than a few lunar masses).
Analyses of timecales for grain dispersal have led to the conclusion
that the dust is maintained by 
replenishment from asteroid collisions and cometary passages.
The evolutionary transition between massive protoplanetary
disks and tenuous debris disks appears to take place early
in a star's life, perhaps within the first 10 Myr (Strom et al. 1989).
However, the timescales and synchronization 
of dispersal processes are largely unknown.

Unprecedented sensitivity at critical wavelengths will enable
\sirtf\ to address key questions in the early evolution of circumstellar 
disks. Weak-line T Tauri stars 
(wTTs) represent a strategically important sample for this purpose.  
These young pre-main sequence objects 
are characterized primarily by reduced signatures of protostellar
accretion. In addition, they lack evidence for outflows and 
display diminished or absent near-infrared excess emission in contrast 
with their cTTS counterparts. 
IRAS and ISO detected far-infrared excesses in only a small fraction of wTTs,
but these early efforts lacked the sensitivity needed 
to detect the masses of material associated with nearby
debris disks at the distances of the nearest star-forming clouds. 
Consequently, it remains unclear whether wTTs are the evolutionary 
descendants of classical T Tauri stars or simply represent a
population of stars {\it without} disks altogether.  
\sirtf\ studies of the wTTs population will bridge this gap in 
understanding and address the following key scientific questions:  

1.  What is the frequency of debris 
disks in the low-mass stellar population at ages of less than 5 Myr?  
Our program will establish the general presence or absence of 
disks in the wTTs sample together with the evolutionary
implications of either possibility.  

2.   On what timescale does the metamorphosis from primordial to
debris disk occur?
By comparing cTTs and wTTs disk properties with those of the disks 
around young main-sequence stars in clusters, we can infer their
evolutionary progress and synchronization.  For example,
grain-growth models predict a rapid transition
from the opaque to translucent states, but the replenishment of
small grains by planetesimal collisions may lead to a longer
transition.  The dispersion in disk properties within a coeval 
sample of stars will also provide a sensitive indicator of the diversity 
of evolutionary pathways for a forming planetary system.

3.   Do most disks develop inner holes or gaps
during the dissipation process?
This aspect of disk structural evolution is observable 
through the signature of the \sed, and our observations will
indicate whether disk dissipation and planetesimal formation proceeds at
an equal pace throughout the disk, or at an accelerated pace in the
inner disk region.

\subsection{Sample Selection}

The selection of wTTs targets was driven by three considerations.  First,
targets were sought that were associated with the large molecular clouds
being mapped.  The proper motion dispersion for wTTs at 140 pc distance
is sufficient for the stars to move several degrees away from their natal
clouds on a timescale of 10 Myr (Hartmann et al.\  1991), so targets were
restricted to a zone within $5\degree$ from the cloud boundaries.
Second, nearby wTTs targets were chosen because they allow the best
sensitivity to dust excess.  For this reason, only targets associated with
clouds closer than 160 pc were considered.  Third, we required that 
stars show evidence of young age: high levels of chromospheric activity as 
traced by X-ray emission detected by ROSAT, and strong Li I 6707 \AA\ 
absorption at levels exceeding those of Pleiades stars of the same
spectral type.

On the basis of  
these criteria, we selected 190 targets associated with the Chamaeleon, 
Lupus, Ophiuchus, and Taurus clouds\footnote{Although the Taurus molecular
cloud is not mapped by c2d, 
parts of the region are being mapped by \sirtf\ Guaranteed Time 
Observers and will be available for comparison.}.  These are distributed
as follows: 30 stars from Chamaeleon 
(Covino et al. 1997), 60 from Lupus (Wichmann 
et al. 1999), 40 from Ophiuchus (Mart\'{\i}n et al. 1998), and 60 from Taurus
(Wichmann et al. 2000).  A few additional stars were selected from
the catalog of 
Herbig \& Bell (1988).  Spectral types in the overall sample range from 
G5 to M5.  Thirty of the targets fall within the boundaries of the large 
cloud maps, leaving 160 to be measured separately.  With the observing
strategy outlined below, these 160 targets require
50 hours.

\subsection{Planned Observations and Expected Results}

We will search for excess infrared emission in our wTTs sample
by means of photometric measurements at wavelengths\footnote
{While highly desirable, 160 \micron\ photometry of our target sample is
impractical due to the high sky backgrounds in the regions surrounding
molecular clouds.}  
of 3.6, 4.5, 5.8, 8.0, 24, and 70 $\mu$m.
At each target, single $5\arcmin \times 5\arcmin$ fields
will be imaged with the IRAC and MIPS cameras.  The IRAC observations
consist of a single 0.6 second and two 12 second exposures. This sequence 
is designed to detect the stellar photospheres at S/N $>$ 50 in 
all four IRAC bands.  At 24 $\mu$m, the MIPS
observation strategy will detect the stellar photosphere at
a S/N level of 20.  For the typical target, 2 photometry cycles
with a 3 second exposure time are required to reach this sensitivity;
actual exposure time will be tailored to the expected photospheric
brightness of each star.  At 70 $\mu$m, robust detection of a stellar
photosphere would be prohibitively time-consuming and probably
impossible due to cirrus and extragalactic confusion.  Consequently,
we have aimed to achieve a more modest goal: S/N level greater than 5
for the detection of a $\beta$ Pictoris-like debris disk 
characterized by a fractional luminosity of $\sim$ 0.001 and 
70 $\mu$m flux density ten times the
stellar photosphere.  For the typical target,
2 photometry cycles with the 10 sec exposure time are needed to achieve
this. As at 24 $\mu$m, the planned exposure time is tailored to the
brightness of the individual star.

\begin{figure}
\plotfiddle{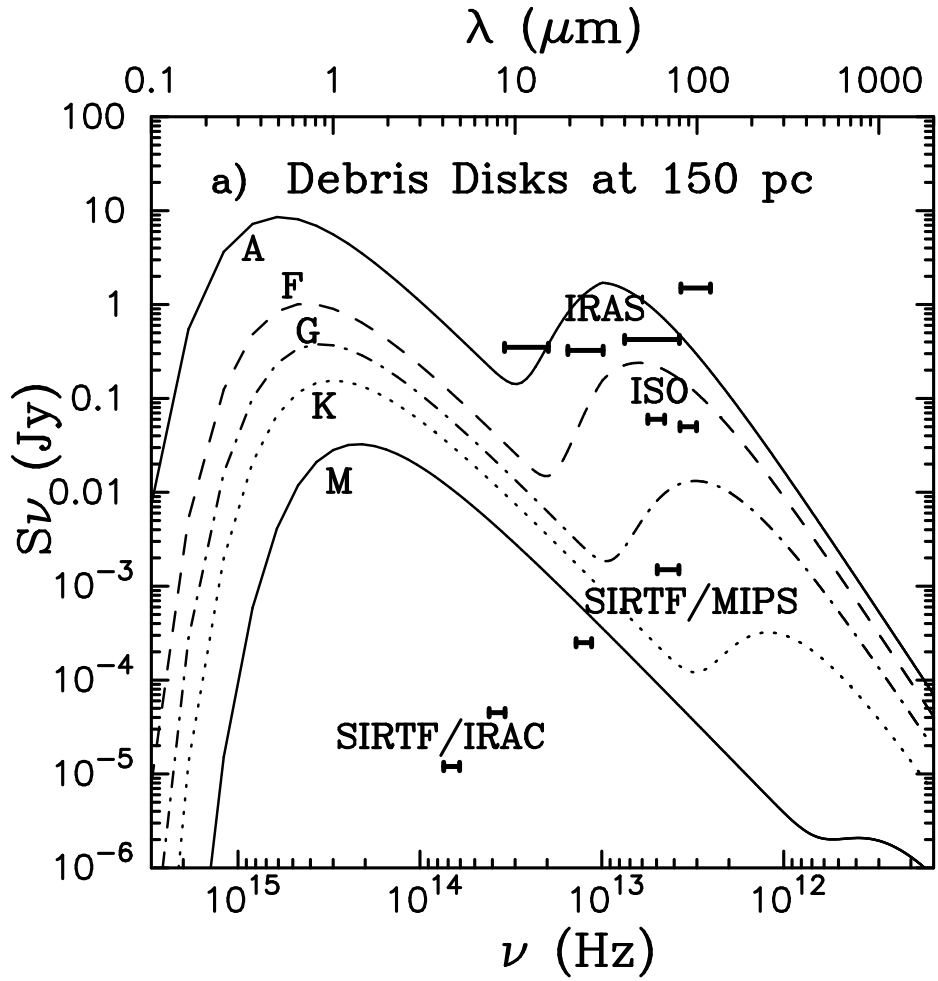}{6.5in}{-90}{65}{65}{-250}{450}
\figcaption{Flux density distributions for model debris disks around stars of
various spectral types in nearby star-forming regions. Model properties
are based on observations of debris disks around A stars, with 0.1
\mmoon\ of 30 $\mu$m-sized dust grains in a zone extending from 30 to 60
AU radius. Horizontal bars mark $5 \sigma$ sensitivity levels for our
survey and for IRAS and ISO. 
\label{debris}
}
\end{figure}

The presence of infrared excess in an individal source can be readily
determined by comparing its colors to those of \sirtf\ standard stars.
To measure absolute levels of the excess, we will compare the observed
flux densities to those of model photospheres appropriate to each
object's spectral type, normalized to match available 2MASS
photometry.  Figure \ref{debris} shows how the survey sensitivity compares to
the emission from various stars surrounded by a debris disk.  
The survey will readily detect disks like
that around $\beta$ Pic, even if they have evacuated inner holes with 
radii as large as 30 AU.  Disks an order of magnitude more tenuous will 
still be detected if their inner edge is at a radius of 5 AU.  Fitting of
model spectral energy distributions to the  \sirtf\ measurements will allow
the disk optical depth, inner radius, and radial density/temperature
profiles to be constrained. 

\subsection{Ancillary and Complementary Data} \label{starsanc}

Two ground-based observing programs are being carried out in support of
the c2d weak-line T Tauri star study.  The first is ancillary echelle
spectroscopy using the 4-m telescopes at KPNO and CTIO.  These spectra
provide uniform measurements of spectral type, metallicity,
lithium abundance, and chromospheric activity in the entire target 
sample, directly supporting the determinations of infrared excess and 
stellar age.  The second is an adaptive optics (AO) search for source 
multiplicity using the ESO Adonis instrument (Hogerheijde et al. 2003).
The presence of companion stars has been shown to have a strong effect on the
frequency of circumstellar disks (Jensen, Mathieu, and Fuller 1996);
by understanding multiplicity in our sample, its effect on the disk
properties can be isolated.  


\section{The Evolution of the Building Blocks of Planets} \label{irs}

\subsection{Scientific Questions}

Spectroscopy complements imaging and is essential to understand the
physical and chemical state and evolution of gas and dust surrounding
young stellar objects.  The mid-infrared wavelength range encompassed
by \sirtf\ is particularly rich in diagnostic features, each of which
probes different aspects of the star- and planet-formation process (see
Table \ref{tabspec}). Indeed, bands of solid-state material and the fundamental
lines of the most abundant molecule, H$_2$, can be studied {\it only}
in the mid-infrared.  In the 75 hour c2d \irs\ program, high
signal-to-noise spectra will be obtained over the full 5--40 $\mu$m range 
[high resolution ($R\approx 600$) over the 10--37 \micron\ range]
for all phases of star- and
planet-formation up to ages of $\sim$5 Myr for at least 170 sources.
The MIPS-SED mode at 50--100 $\mu$m will also be used in the second
year of the program to characterize the longer wavelength silicate and
ice features of a disk sub-sample. Previous spectroscopic studies, 
e.g., with \iso, only had the sensitivity to probe high- or
intermediate-mass young stellar objects.  \sirtf\ will permit the first
comprehensive mid-infrared spectroscopic survey of solar-type young
stars.

\begin{figure*}
\plotfiddle{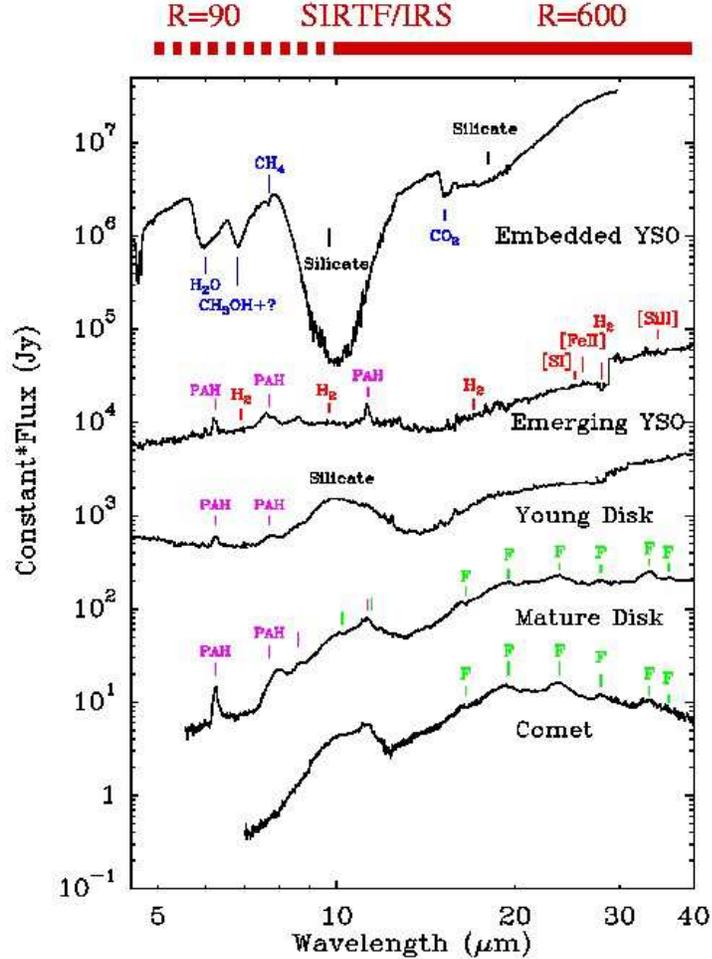}{5.2in}{0}{60}{60}{-200}{-40}
\figcaption{\label{irsfig}
ISO-SWS mid-infrared spectra of newly-formed stars in different
stages of formation. From top to bottom -- in a rough evolutionary
sequence -- the spectra change from dominated by solid-state absorption
features (ices and
amorphous silicates) to gas emission lines and PAH features to amorphous
silicate emission and eventually crystalline silicate features (labeled
``F''). Note the similarity of the spectra of mature disks with that of comet
Hale-Bopp. ISO was able to obtain such spectra only for massive young stars;
SIRTF has the capability to study these features for sun-like objects. Based
on van den Ancker et al. (2000a,b), Gibb et al. 2000, Crovisier et al. 1997,
 and Malfait et al. (1998).
}

\end{figure*}

The \irs\ spectra can be used to address the following questions: 

1.  What do spectroscopic diagnostics tell us about the different
stages of early stellar evolution?  As chronicled in Table \ref{tabspec} and
Figure \ref{irsfig}, there are many mid-infrared spectroscopic features that
are potentially diagnostic for distinct physical and chemical states of
young circumstellar environments.  A classification based on these
features would complement the current classification scheme 
(Lada \& Wilking 1984) based on spectral energy distributions.  
Prime diagnostics in
the earliest embedded stages are the solid CO$_2$ bending mode at 15
$\mu$m, the [S I] 25 $\mu$m line and other atomic lines, the PAH
features, and the H$_2$ lines.  The shape of the 15 $\mu$m solid CO$_2$
band seen in absorption toward protostars is particularly sensitive to
the thermal history of the envelope, and changes in its profile have
allowed young high-mass protostars to be put in an evolutionary
sequence (Gerakines et al.\ 1999). The [S I] 25 \micron\ and H$_2$
lines probe the presence and properties of shocks, whereas the PAH
features signal the importance of ultraviolet radiation.  Together,
they allow us to trace the relative importance of these two processes 
for clearing the envelope.  A particularly exciting prospect is
mid-infrared spectroscopy of disks as they are being unveiled.

In the later evolutionary stages, changes in silicate features become
the main diagnostics.  Indeed, one of the major results from the SWS
instrument on \iso\ is that solid-state evolution of grain minerals
occurs in disks around pre-main sequence stars (Meeus et al.\ 2001).
Some objects show only amorphous dust emission, whereas others have
clear signatures of crystalline silicates and/or PAHs. These variations
may be related to the processes of grain coagulation, settling and
destruction in the circumstellar disk, i.e., its evolution.

2. How does the chemical composition of dust and ices change from
molecular clouds to planetary bodies?  In the embedded phase a
reservoir of volatile and solid materials is delivered to the disk
from the surrounding cloud core.  The composition of the ices in the
embedded phase can be probed by absorption spectroscopy, whereas that
of the silicates in the pre-main sequence phase requires emission
studies.  The low \irs\ resolving power below 10 $\mu$m does not allow
detection of minor ice species, but the major ice components can be
traced.  Many ices may survive the accretion shock
and will be incorporated into icy grain mantles in the outer reaches
of planetary systems, perhaps to be delivered to inner planets intact.
Indeed, remarkable similarities are found between the composition of
interstellar ices and comets and between the crystalline silicates and
PAHs seen in some disks and those in comet Hale-Bopp (Malfait et al. 1998).
An inventory of the material present at each stage of low-mass envelope
and disk evolution as a function of stellar luminosity, mass, and environment
will be a major goal of the spectroscopy program.  The resulting
spectra will form a powerful data base to compare with spectra of
Kuiper-Belt objects, comets and asteroids obtained in other \sirtf\
programs and to clarify the links between interstellar, circumstellar, and
solar-system material.


3. How does the size distribution of dust grains evolve in
circumstellar environments?  Observational and theoretical evidence
points to a great deal of activity in dust grain evolution during the
gas-rich disk phase, suggesting that grain growth to a
km-sized population of planetesimals may occur at this stage. The
silicate emission widely observed from gas-rich disks arises from
small ($\sim$0.5 $\mu$m) grains, while the underlying continuum stems
from larger particles.  Further, the overall excess shortward of 200
$\mu$m scales approximately as the ratio of the disk (dust)
photosphere to the gas scale height (Chiang et al. 2001).  
The low noise \irs\ spectra automatically cover the lowest
pure rotational emission lines of H$_2$ at 28 and 17 $\mu$m, which
may be detectable if the gas and dust temperatures are sufficiently
different. Successful detection of these lines would permit an
independent assessment of the dust coagulation and gas dissipation
time scales, processes of great importance for planet formation.

It is important to note that the spectral energy distributions
(SEDs) themselves do not give a unique answer to grain growth
because of the degeneracy between disk mass and grain size
distribution in radiative transfer models, especially if the
physical size of the disk is uncertain (Chiang et al. 2001).
However, the combination of SEDs with complementary spatially
resolved infrared and millimeter-wave images of disks does resolve
the parameter correlation, making it possible to constrain the
relative settling of the dust versus the gas in gas-rich disks
and the earliest stages of planetesimal formation up to roughly
millimeter sizes.  Our c2d program will provide hundreds of
objects for more detailed follow-up.

4. What is the spectral evolution of substellar objects?  
Do they form as stars do, or do they form as companions to stars? 
A clue that they form as stars do is the existence of a disk
around many young substellar objects
(Natta \& Testi 2001; Apai et al. 2002; Liu et al. 2003). 
Another important question is whether their atmospheres are dusty at later
times.  Atmospheric spectra are
predicted to show significant changes with age and mass (e.g., Burrows
et al.\ 1997).  Apart from limited photometry, nearly nothing is known
observationally about the mid-infrared spectra of brown dwarfs, young
or old.  The young brown dwarfs and super-Jupiters discovered in the
IRAC and MIPS surveys ($\sim$ 100 expected) therefore form a critical
second look population for study with the \irs.

\subsection{Sample Selection}

\begin{figure}
\plotfiddle{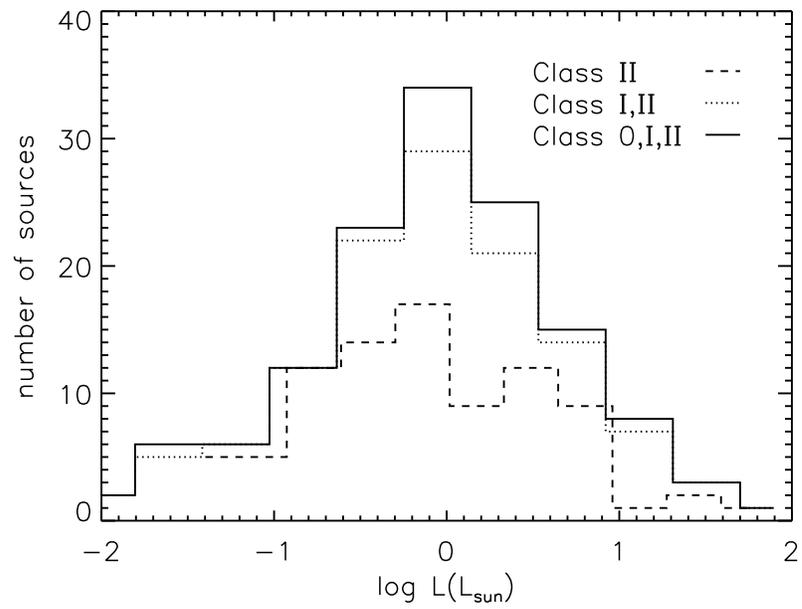}{5.0in}{90}{50}{50}{200}{0}
\figcaption{Distribution of luminosity over the sample for the first-look IRS
observations. Stellar luminosities are calculated from fits to SEDs compiled 
from the literature (about 80\% of sample is included in this plot).
\label{irslum}
}
\end{figure}

\begin{figure}
\plotfiddle{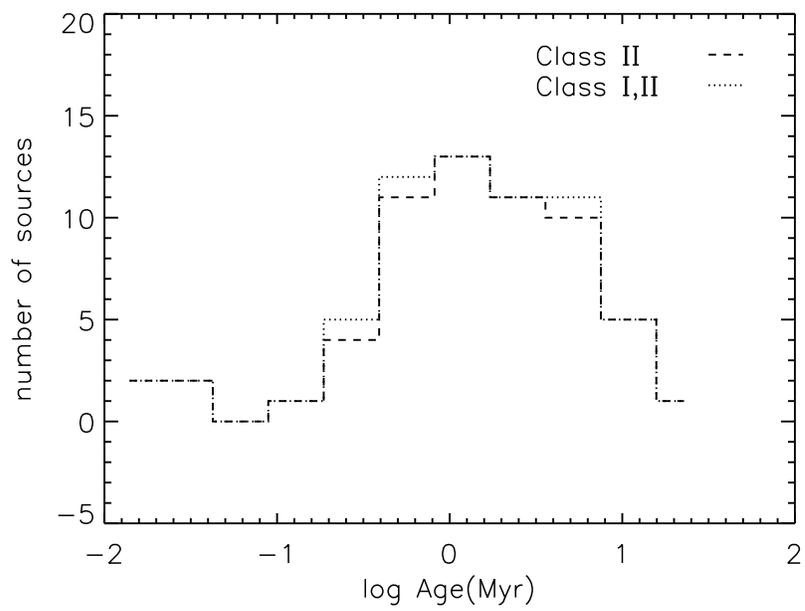}{5.0in}{90}{50}{50}{200}{0}
\figcaption{Distribution of ages over the sample for the first-look IRS
observations. Note that age information is quite incomplete and only
one-quarter of the sample is represented in this plot.
\label{irsage}
}
\end{figure}

The \irs\ observations are divided into two sets with roughly equal
time, the first being observations of known embedded and
pre-main-sequence stars and the second consisting of follow-up
spectroscopy of sources discovered in the IRAC and MIPS mapping
surveys.  The source list for the first-look program was restricted
primarily to low-mass young stars, defined as having masses $M\lesssim
2\rm ~M_{\odot}$ with ages younger than $\sim$5 Myr, for
minimal overlap with existing infrared spectroscopy.
Within these criteria the selection contains a broad representative
sample of young stars with ages down to 0.01 Myr and masses down to
the hydrogen burning limit or even less, if possible.
Figure \ref{irslum} shows the distribution of sources over
luminosity, and Figure \ref{irsage} shows the distribution
over age.

The initial selection of sources to be observed with the \irs\ was
constrained to the cloud regions scheduled to be mapped by IRAC.  A
large list of Class 0, I and II sources was compiled, primarily from
the existing near- and mid-infrared surveys of the mapped clouds
(Persi et al.\ 2000, Bontemps et al.\ 2001).
All overlaps with \gto\ high spectral resolution \irs\ targets were
removed.  The SIMBAD database was searched for additional observations
and modeling of every source remaining in the list. All available
photometric points were collected and typical parameters such as
bolometric luminosity, SED Class, effective temperature, and optical
extinction were entered into a concise database from which the final
selection could be made.  Stars with no measured mid-infrared fluxes
were dropped first. The availability of sensitive ISOCAM surveys
covering part of the mapped regions ensured that the fainter end of
the luminosity function is well represented.  

Most sources with $\rm 5-25~\mu m$ fluxes less than $\rm 200~mJy$ were
also discarded in the first look \irs\ sample due to time limitations;
some fainter sources of special interest, such as a few stars
known to have an edge-on disk, were retained. Stellar ages of the
Class II sources were calculated by fitting the extinction-corrected
temperature and bolometric luminosity from the database to the
evolutionary tracks by Siess et al. (2000), assuming solar abundances.  Stars
with a calculated age of more than 5 Myr were subsequently discarded
since systems of this age and older are covered in the \feps\
Legacy spectrophotometry program, albeit largely at low spectral
resolution. The flux limit to be achieved in the first-look survey
corresponds to a typical source luminosity of $\rm 0.1~L_{\odot}$
and mass of $\rm 0.1~M_{\odot}$ in the Ophiuchus, Chameleon, Serpens
and Lupus clouds at an age of 1 Myr.

Care was taken at this point to verify that all stellar ages and
masses within the limits of the full database were represented in the
final sample. For completeness and comparison purposes, a few sources
were added such as background stars and certain Herbig Ae stars that
were well characterized by the ISO-SWS.  The final source list for the
first-look survey consists of about 170 unique targets, of which about
10\% belong to Class 0, 20\% to Class I and the rest to Class II.  A
roughly equal amount of time is reserved to the spectroscopic
follow-up of interesting sources found in the imaging and photometry
surveys, complementary MIPS-SED observations of selected first-look
\irs\ sources, and spectral mapping of one cloud core.

\subsection{Planned Observations and Expected Results}

All first-look targets will be observed using the \irs\ staring
mode in each of its four modules, except for those sources that are
part of various \gto\ programs involving the low resolution modules.
For those stars, only the high resolution 10-37 $\mu$m spectra will
be acquired as part of the c2d \irs\  effort. Whenever possible,
cluster observations will be used to reduce the slewing and peak-up
overheads. Specifically, the moderate precision cluster mode
will be used in conjunction with the blue filter on the peak-up array.

The integration times for the short-high and long-high modules were
fixed such that theoretical S/N ratios of at least 100 and 50 are
obtained for sources brighter and fainter than $\rm 500~mJy$,
respectively. The spectra to be taken using the short-low modules
always reach theoretical S/N ratios of $>$100.  In contrast to the
scheduled \gto\ observations of large numbers of young stars, typically
with the low resolution \irs\  modules, the c2d \irs\ program focuses on
long integration times in the high resolution modules, ensuring high
dynamic range even on weak sources.  Instrumental fringing may limit
these S/N ratios, so the c2d \irs\ team is leading the development
of defringing tools for \sirtf, as described below.  In order to
evaluate the pointing performance of the spacecraft and assess the
defringing software before beginning the full survey, the c2d \irs\
validation program will include cluster mode observations of stars
with widely varying brightness.

While most of the objects will be observed in the \irs\ stare mode only,
the northwestern Serpens molecular core will be imaged over more than
4 square arcminutes to a $1\sigma$ sensitivity of 2 mJy using the low
resolution \irs\ spectral mapping mode.  This core contains several 
deeply-embedded sources and possesses a complex physical structure with infall,
outflow, and formation of the envelope and disk all occurring within 
30\arcsec\ to 60\arcsec\ ($\sim 0.05-0.1$~pc) of the central star.  Because the
continuum is weak off-source, emission lines can be detected even at
$R=60-120$, and every \sirtf\ pixel may have an interesting spectrum,
ranging from those characterized by deep ice absorption bands toward
the protostars themselves to silicate emission from nearby disk
sources or strong ionized lines at the heads of shocks.  At some
positions, the ``continuum'' emission may even be entirely due to
lines or PAHs, which could significantly affect the interpretation and
classification of objects.

The targeted dynamic range of 50--100 on the continuum in the \irs\
stare observations is driven by the scientific questions outlined
above, especially the desire to study the thermal history of the
envelope through the 15 $\mu$m CO$_2$ bending mode profiles 
and to search for gas phase emission and absorption
features in all phases of star formation. Even at $R=600$, however,
minor grain mantle components will be difficult to detect, and little
or no kinematic information will be gleaned from the gas-phase lines.
In addition, no information will be available on features shortward of
5 $\mu$m.

\subsection{Complementary Data}\label{irscomp}

As a complementary ground-based program, we have therefore initiated
flux-limited surveys in the atmospheric L-band and M-band windows using
the VLT-ISAAC and Keck NIRSPEC (McLean et al. 1998) instruments
in order to examine the fundamental stretching vibrations of the
H$_2$O and CO molecules, respectively. The excellent sensitivity of
these spectrometers has enabled the first high spectral resolution
($R=10,000-25,000$) CO observations of low-mass protostars, revealing
both gaseous and solid CO (Pontoppidan et al. 2003).  
Embedded sources in Taurus and Ophiuchus
show a blend of gas-phase CO absorption and emission profiles that are
related to the simultaneous infall and outflow velocity fields in
protostars (see Pontoppidan et al.\ 2002).  Studies of edge-on disk
sources have been particularly revealing of the solid state
components and velocity fields in accretion disk surface layers and
mid-plane (Boogert, Hogerheijde, \& Blake 2002, Thi et al.\ 2002). In older
Class II disk systems, particularly Herbig Ae stars, emission from CO
and atomic hydrogen lines is observed and likely arises from the
inner disk in or near the dust sublimation radius (see also Dullemond
et al. 2001, Brittain \& Rettig 2002).

A subset of IRS targets with known infrared ice absorption
features is being characterized, using the CSO and OVRO facilities.
Observations of \cooo\ and \hcop\ lines as well as millimeter
continuum emission are used to derive the distribution and physical
conditions (temperature, density, and column density) 
of material along the line of sight.
This information is crucial in locating the ices and understanding
ice evolution indicators (ice column, absorption band profile),
as well as in properly interpreting the spectral energy distribution
(Boogert et al. 2002).

\subsection{Analysis Tools }

The c2d-\irs\ team will enhance the \irs\ pipeline data delivered by the
\ssc\ in several ways. The most important improvement will be in the
defringing of the IRS spectra, because laboratory experiments indeed
show the presence of fringes.  Special software derived from the ISO-SWS
experience is being written for defringing both 1-D and 2-D spectra
(Lahuis \& Boogert 2002). The in-flight fringe characteristics (complexity
and amplitude) are presently unknown, however, which is one of the reasons
why two independent defringing approaches, each with its own merits,
are being exploited.
The first uses a robust method of iteratively fitting sine functions;
the second algorithm minimizes fringe residuals by correcting the flat
field to best match the data. In parallel, a ``complete'' fringe model
is being developed, applying basic optical theory and incorporating
the geometry and optical properties of the \irs\ detectors.  The
routines are written in IDL and the resulting modules are compatible
with the SMART package, developed by the \irs\ instrument team for their
data reduction and management 
(see {\tt http://www.astro.cornell.edu/SIRTF/}). The 1-D pre-launch
modules have been delivered to the \ssc\ for use by the community. Other
improvements to the pipeline data will come from monitoring of the
data quality and calibration over time.


\section{Data Products} \label{datprod}

The source samples and observations are summarized in Table \ref{summarytab}.
The data products (summarized in Table \ref{productsum})
include the following categories of observations and 
analyses: \sirtf\ mapping observations together with an associated
source catalog and ancillary mapping of the same
regions at \mm\ wavelengths; \sirtf\ photometry and NOAO
optical spectroscopy of associated weak-line T Tauri stars; and IRS
spectroscopy of a sample of objects at all stages of early evolution.
These data can be used by the community to address far  
more than just the principal science areas listed above. We 
describe the particulars of these products below in hopes
that other researchers will easily recognize their utility across 
a broader range of scientific studies. \sirtf\ and ancillary products will be 
made available to the SIRTF Science Center for community access through 
the Infrared Sky Archive (IRSA). 
Because the complementary data are being obtained by many researchers
operating under many different guidelines about archiving, the availability
of those data will be handled on a case by case basis.
We hope to make most of the complementary data accessible 
eventually, either through IRSA or through our web sites.

\subsection{Clouds and Cores}

We will examine and, where necessary, enhance the basic calibrated data 
(BCD) delivered by the pipeline at the SIRTF Science Center.
We will provide to IRSA the enhanced calibrated data (ECD) for all the fields
incorporated in our IRAC and MIPS imaging studies of nearby star-forming 
regions. In addition, we will provide mosaicked maps of the clouds and cores; 
these products are referred to in Table \ref{productsum} as mosaics.
A band-merged catalog of sources contained within these images will
also be compiled and provided to IRSA. The
catalog will include cross-identifications with known pre-main 
sequence objects, foreground stars, and transient sources (for example,
asteroids in the field of view). This resource should allow the community easy
access to \sirtf\ photometry of any individual objects that fall within
our field of view, as well as the capability to carry out 
statistical studies of color- or magnitude-selected samples
with the help of the IRSA search engine. 
As noted in Table \ref{productsum}, all of our large clouds except
Cham II include ``cut-outs," regions observed by \gto s. Because data
on these regions will not be available to us until 12 months after they are
observed, delivery of full mosaics and catalogs for those clouds will
depend on the observation date of those cut-outs.

\subsection{Data Products for weak-line T Tauri Stars}

IRAC and MIPS images of weak-line T Tauri stars will be processed
in the same manner as the images of isolated cloud cores and extended
molecular clouds and the results provided to IRSA.  
The c2d team will also deliver the following: 1) enhanced calibrated data
(ECD) consisting of images in six \sirtf\ bands; 2) a catalog of the sample
which includes cross-identifications with observations in the literature
and spectral energy distributions including both \sirtf\ and published
measurements; 3) new high-resolution optical spectra for the sample, 
together with derivative properties, such as equivalent widths of 
lithium and H$\alpha$, and with radial velocities and identification 
of any double-lined spectroscopic binaries; and 4) a list of positions and 
fluxes for serendipitous sources. Archival researchers will be able to 
access \sirtf\ photometry, band-merged with other relevant photometric
measurements for any individual source of interest, or to view selected
properties of the entire sample with the help of IRSA.

\subsection{IRS Data Products}

The data products will consist of best-effort reduced and
defringed spectra, together with an identification of the spectral
features where possible. 
Results from our complementary VLT, Keck, and CSO/SEST
programs will be made available through publications in refereed
journals. The reduced spectra will appear on the c2d web site at the
time of the final data delivery from our program.

The most important improvement the c2d-IRS team will make
to the IRS pipeline data is defringing of the spectra. Special
software has been written for this purpose (Section 4.4).  Other
improvements to the pipeline data will come from monitoring the data
quality and calibration over time.
The 1-D pre-launch
modules have been delivered to the SSC for use by the community. Other
improvements to the pipeline data will come from monitoring of the
data quality and calibration over time.

\subsection{Modeling Tools}\label{modeltools}

Simple modeling tools
that fit the photometric data of individual sources with 
emission from protostellar cores and circumstellar disks will
be available to help identify the nature of catalogued sources
({\tt http://wits.ipac.caltech.edu/}).
Simple analysis programs for modeling SEDs from disks are being developed
for our team based on the formalism by Dullemond et al.\ (2001) and
will be made available through the c2d website.


\section{Summary} \label{summary}

The c2d program will provide a legacy for future research on star and
planet formation. By selecting samples with attention to coverage of
the relevant parameters, we hope to provide a data base for unbiased
statistical studies of the formation of stars and substellar objects.
Ancillary and complementary data from other wavelength regimes will
complete the picture. Analysis and modeling tools will assist researchers
in getting the most out of the data base. 
The source lists may be found on the c2d web page, noted in the
introduction, by following the link to ``source list." Table \ref{summarytab}
summarizes the samples (columns) and observations (rows) to be collected, 
and Table
\ref{productsum} lists the data products and anticipated delivery
dates. Of course, source lists, observations, data products, and delivery
dates may be modified if in-flight performance differs from what was predicted.

We anticipate extensive 
follow-up studies of these samples with \sirtf\ itself and with future 
missions such as \SIM, \herschel, \sofia, \ngst, and \tpf, as well as 
with ground-based instruments such as \sma, \carma, and \alma.

\acknowledgements

This research has made use of NASA's Astrophysics Data System,
the SIMBAD database, operated at CDS, Strasbourg, France,
and the NASA/ IPAC Infrared Science Archive, which is operated by the 
Jet Propulsion Laboratory, California Institute of Technology, 
under contract with the National Aeronautics and Space Administration.
We thank L. Cambr\'esy and P. Padoan for supplying electronic
versions of data.
This material is based upon work supported by the National Aeronautics
and Space Administration under Contract No. 1224608 issued
by the Jet Propulsion Laboratory.
The Leiden SIRTF legacy team is supported by a Spinoza grant from the
Netherlands Foundation for Scientific Research (NWO) and by a grant from
the Netherlands Research School for Astronomy (NOVA). 
CWL is partially supported by grant R01-2000-000-00025-0
from the Basic Research Program of the Korea Science and Engineering
Foundation.




\begin{deluxetable}{rrrr}
\tablecolumns{4}
\tablecaption{Clouds to be Surveyed \label{cloudtab}}
\tablewidth{0pt}
\tablehead{
\colhead{Cloud}	&
\colhead{Distance}	&
\colhead{Area\tablenotemark{a} } &
\colhead{Time\tablenotemark{b}}	\\
\colhead{}	&
\colhead{(pc)}	&
\colhead{(sq. deg.)} &
\colhead{(hours)}
}

\startdata
Perseus		&	320	&	3.8	&	52.7 \\
Ophiuchus	&	125	&	8.0	& 	78.1 \\
Lupus		&	125	&	2.4	&	46.6 \\
Serpens		&	310	&	0.8	&	12.9 \\
Chamaeleon II	&	200	&	1.1	&	15.2 \\

\enddata
\tablenotetext{a}{Area mapped with IRAC; MIPS will cover a larger area.}
\tablenotetext{b}{Time for full maps with IRAC and MIPS, including off-cloud
comparison fields, using SPOT6.2.}
\end{deluxetable}

 
\begin{deluxetable}{l|rrrrr}
\tablecolumns{6}
\tablecaption{Expected Stellar Background 
Count\label{background_tab}\tablenotemark{a}}
\tablewidth{0pt}
\tablehead{
\colhead{$\lambda$}     &
\colhead{Cham II}      & 
\colhead{Lupus}        & 
\colhead{Ophiuchus}   &
\colhead{Perseus}   &
\colhead{Serpens}
}
\startdata
L(3.6 $\mu$m)    & 400 & 7000 & 900  & 110 & $10^4$   \\
N(10.2 $\mu$m)   &  20 &  300 &  45  &  10 & 600  \\
25 $\mu$m        &   1 &    8 &   1  &   1 & 15   \\
\enddata
\tablenotetext{a}{
Expected background counts are from the model of Wainscoat et al. (1992)
and are for a $5\arcmin\times5\arcmin$ observing area. The wavelength
bands do not exactly correspond to the \sirtf\ bands.}
\end{deluxetable}


\begin{deluxetable}{rrrrr}
\tablecolumns{5}
\tablecaption{Sensitivities of Cloud Survey \label{senstab}}
\tablewidth{0pt}
\tablehead{
\colhead{$\lambda$}	&
\colhead{Sensitivity\tablenotemark{a}}	&
\colhead{Sensitivity\tablenotemark{b}}	&
\colhead{Saturation }	&
\colhead{Saturation }	\\
\colhead{($\micron$)}	&
\colhead{(mJy)}	&
\colhead{(mag)}	&
\colhead{(Jy)}	&
\colhead{(mag)}	
}

\startdata
3.6	&	0.015	&	18.0	&	0.040  &	9.4	\\
4.5	&	0.019	&	17.3	&	0.031  &	9.3	\\
5.8	&	0.060	& 	15.6	&	0.093  &	7.6	\\
8.0	&	0.083	&	14.6	&	0.077  &	7.1	\\
24	&	0.83	&	9.8	&	0.206  &	3.9 	\\
70	&	5.2	&	5.7	&	0.258  &	1.5     \\

\enddata
\tablenotetext{a}{Sensitivities are 3 $\sigma$ for 24 sec total time.}
\tablenotetext{b}{Magnitudes based on power-law interpolation and extrapolation
of zero-point fluxes from standard bands to the \sirtf\ wavelengths.}
\end{deluxetable}

\begin{table*}
\caption{Selected mid-infrared spectral features \label{tabspec}}
\footnotesize
\begin{tabular}{rllrll}
\tableline
  $\lambda$ \ \  & Species & Diagnostic 
  & $\lambda$\ \  & Species & Diagnostic \\
   (\micron)   &         & &  (\micron)   & & \\
\tableline
6.0 & H$_2$O ice &            Bulk of ice &
   12.8 & [Ne II]             & Radiation field, shocks \\
6.2 & PAH       & UV radiation, carbon.\ material 
 & 15.2 & CO$_2$ ice          & Thermal history \\
6.8 & Unid. ice      & Processed ices (UV/cosmic ray) 
 & 15.6 & [Ne III]           & Radiation field \\
6.9 & H$_2$ S(5)           & Photon vs shock heating 
 & 17.0 & H$_2$ S(1)         & Mass and T warm gas \\
7.7 & CH$_4$ ice        & Building organics, solar system 
 & 18.5 & (Mg,Fe)SiO$_3$     & Cryst. pyroxenes \\
7.7 & PAH            & UV radiation, carbon.\ material 
   & 23.0   & FeO                 & Oxides \\
8.0 & H$_2$ S(4)      & Photon- vs shock-heating  
  & 25.2 & [S I]                & Shocks  \\
8.6 &  PAH & UV radiation, carbon.\ material &
  27.5 & Mg$_2$SiO$_4$        & Cryst. silicates, heating \\
9.7 & Amorp. sil.\    & Bulk of dust &
   28.2 & H$_2$ S(0)            & Mass and T warm gas \\
9.7 & H$_2$ S(3)            & Photon vs shock heating 
 & 33.5 & Mg$_2$SiO$_4$         & Cryst. silicates (enstatite) \\
11.3 & Mg$_2$SiO$_4$      & Cryst. silicates, heating 
& 35.8 & Mg$_2$SiO$_3$         & Cryst. pyroxenes \\
11.3 & PAH                 & UV radiation, carbon. material 
  & 61  &  cryst. H$_2$O        &  Cryst. ices, heating \\
12.2 & H$_2$ S(2)          & T warm gas, ortho/para ratio 
  & 70  &  cryst. sil.        &  Cryst. silicates, heating \\
\tableline
\end{tabular}
\end{table*}


\begin{deluxetable}{l|rrrrl}
\tablecolumns{6}
\tablecaption{Summary of Samples and Observations\label{summarytab}}
\tablewidth{0pt}
\tablehead{
\colhead{Item}     &
\colhead{Clouds}      & 
\colhead{Cores}        & 
\colhead{Disks}   &
\colhead{IRS Targets}   &
\colhead{Comments}
}
\startdata
Number         & 5      &156     &190       &$170+$   & First Look\tnm{a} \\
Area (sq.deg)  & 16     & 1      &\nodata   &\nodata  &            \\
IRAC           & map    & map    & phot     & \nodata & 3.6--8 \micron\  \\
\ \ Sens. 3.6 \um  &0.015   &0.015   & 0.015    & \nodata & (mJy, 3$\sigma$) \\
\ \ Sens. 4.5 \um  &0.019   &0.019   & 0.019    & \nodata & (mJy, 3$\sigma$) \\
\ \ Sens. 5.8 \um  &0.060   &0.060   & 0.060    & \nodata & (mJy, 3$\sigma$) \\
\ \ Sens. 8.0 \um  &0.083   &0.083   & 0.083    & \nodata & (mJy, 3$\sigma$) \\
MIPS           & map    & map    & phot     & \nodata & 24, 70 \micron\ \\
\ \ Sens. 24 \um   &0.83  &0.53 & varies\tnm{b} & \nodata & (mJy, 3$\sigma$) \\
\ \ Sens. 70 \um   &5.2   &4.7  & varies\tnm{b} & \nodata & (mJy, 3$\sigma$) \\
IRS            &select\tnm{c}  &select\tnm{c}  &\nodata   & L, H\tnm{d} &             \\
\ \ Signal/Noise  &50--100 &50--100 &\nodata   & 50--100 & On continuum   \\
Complementary  & yes    & yes    & yes      & yes     & Ancillary as noted \\
\ \ Visible images &  map  & select\tnm{c} &\nodata   & \nodata & R, i, z images   \\
\ \ Visible spectra &\nodata &\nodata & 190     & \nodata & Ancillary \\
\ \ NIR AO         &\nodata &\nodata & select\tnm{c}   & \nodata & Adaptive optics \\
\ \ MIR spectra    & select\tnm{c} & select\tnm{c} &\nodata   & select\tnm{c}  & \\
\ \ mm/submm\tnm{e}& map    & map    &\nodata   & select\tnm{c}  & Continuum maps \\
\ \ mm interf.\tnm{f} & select\tnm{c} &\nodata &\nodata &select\tnm{c} & Continuum maps \\
\ \ mm spectra     & map    & map    &\nodata   & select\tnm{c}  &  Spectral maps \\
\enddata
\tablenotetext{a}{
A roughly similar number of spectral sources will be observed in 
second look mode, based on the results of the continuum surveys.}
\tablenotetext{b}{
Adjusted to achieve a certain signal/noise depending on the star.}
\tablenotetext{c}{
Selected objects or regions within this sample will be
observed in the indicated mode; the coverage is not complete.
For example, 129 of the IRS targets are toward the large clouds.}
\tablenotetext{d}{
Spectra will be taken with both low resolving power
($R \approx 60-120$) and high resolving power ($R \approx 600$) modules.}
\tablenotetext{e}{
Large-scale maps made with bolometer arrays (Bolocam, SIMBA, SCUBA, MAMBO).
Maps of three clouds (Perseus, Ophiuchus, and Serpens) are ancillary data.}
\tablenotetext{f}{
Small-scale maps of selected regions made with interferometers (BIMA, OVRO).}
\end{deluxetable}


\begin{deluxetable}{lll}
\tablecolumns{3}
\tablecaption{Anticipated Data Products and Delivery Dates\tnm{a}
\label{productsum}}
\tablewidth{0pt}
\tablehead{
\colhead{Date\tnm{b}}     &
\colhead{Product}     &
\colhead{If Observed By\tnm{c}}     
}
\startdata
L + 9  & Sampler: validation observations, all modes       & L + 6  \\
       & Ancillary NOAO optical spectroscopy of wTTs       & \nodata \\
       & Catalog of IRAC/MIPS results for wTTs             & L + 6   \\
L + 15 & Initial band-merged, cross-id  catalog for wTTs   & L + 12   \\
       & Initial band-merged, cross-id catalog for cores   & L + 12   \\
       & ECDs for cores, cloud areas                       & L + 12   \\
       & Spectra, cataloged features, first-look targets   & L + 12   \\
L + 21 & Final band-merged, cross-id catalog for wTTs      & L + 15   \\
       & Final band-merged, cross-id catalog for cores     & L + 15  \\
       & Mosaics for cores                                 & L + 15  \\
       & Mosaic for Cham II                                & L + 15  \\
       & Defringed spectra, cataloged features, first-look & L + 15  \\
L + 27 & Mosaics for Clouds                                & L + 9\tnm{d} \\
       & Final, band-merged, cross-id catalog for clouds   & L + 9\tnm{d} \\
       & Ancillary submm cloud maps                        & L + 9\tnm{d} \\
       & Catalog of small extended sources                 & L + 9\tnm{d} \\
       & Catalog of transient sources                      & L + 9\tnm{d} \\
       & Defringed spectra, cataloged features, second-look & L + 24  \\
L + 31 & Mosaics for any delayed clouds                    & L + 12\tnm{d} \\
       & Updated catalogs for any delayed clouds           & L + 12\tnm{d} \\
       & Defringed spectra, cataloged features, second-look & L + 28  \\
       & Complementary data, where possible                & \nodata  \\
\enddata
\tablenotetext{a}{The products and delivery dates are based on assuming
that the spacecraft and all instruments function normally and that the 
data pipeline runs smoothly. If any of those assumptions are wrong, 
products may be delayed or even eliminated.}
\tablenotetext{b}{The dates are all given in months after Launch.}
\tablenotetext{c}{Products will be available on sources observed by this date.}
\tablenotetext{d}{All large clouds except Cham II have ``cut-outs," areas
observed by \gto s; these areas will not be available to us until 12 months
after they are observed. Delivery of our final cloud images and catalogs
depends on the observation date of these cut-outs.}
\end{deluxetable}

\end{document}